\documentclass[useAMS,usenatbib]{mn2e_personal}
\usepackage{graphicx}

\title[Protoplanetary migration in sub-Keplerian flows]{How initial and boundary conditions affect protoplanetary migration in a turbulent sub-Keplerian accretion disc: 2D non viscous SPH simulations}
\author[V. Costa, V. Pirronello, G. Belvedere, A. Del Popolo, D. Molteni, G. Lanzafame]{V. Costa$^{1}$\thanks{E-mail:
vcosta@oact.inaf.it}, V. Pirronello$^{1}$, G. Belvedere$^{2}$, A. Del Popolo$^{2}$, D. Molteni$^{3}$, G. Lanzafame$^{4}$\\
$^{1}$Dipartimento di Metodologie Chimiche e Fisiche per l'Ingegneria, Univesit\`a di Catania, Viale A. Doria 6, I-95125, Catania, Italy\\
$^{2}$Dipartimento di Fisica e Astronomia, Universit\`a di Catania, Sezione Astrofisica, Via Santa Sofia 78, I-95125, Catania, Italy\\
$^{3}$Dipartimento di Fisica e Tecnologie Relative, Universit\`a di Palermo, Viale delle Scienze, I-90128, Palermo, Italy\\
$^{4}$Osservatorio Astrofisico di Catania, Istituto Nazionale di Astrofisica, Via S. Sofia 78, I-95123, Catania, Italy
}
\begin{document}

\date{}
\pagerange{\pageref{firstpage}--\pageref{lastpage}} \pubyear{2009}

\maketitle

\label{firstpage}

\begin{abstract}

Current theories on planetary formation establish that giant planet formation should be contextual to their quick migration towards the central star due to the protoplanets-disc interactions on a timescale of the order of $10^5$ years, for objects of nearly 10 terrestrial masses.
Such a timescale should be smaller by an order of magnitude than that of gas accretion onto the protoplanet during the hierarchical growing-up of protoplanets by collisions with other minor objects.
These arguments have recently been analysed using N-body and/or fluid-dynamics codes or a mixing of them. 
In this work, inviscid 2D simulations are performed, using the SPH method, to study the migration of one protoplanet, to evaluate the effectiveness of the accretion disc in the protoplanet dragging towards the central star, as a function of the mass of the planet itself, of disc tangential kinematics.
To this purpose, the SPH scheme is considered suitable to study the roles of turbulence, kinematic and boundary conditions, due to its intrinsic advective turbulence, especially in 2D and in 3D codes.
Simulations are performed both in disc sub-Keplerian and in Keplerian kinematic conditions as a parameter study of protoplanetary migration if moderate and consistent deviations from Keplerian Kinematics occur. 
Our results show migration times of a few orbital periods for Earth-like
planets in sub-Keplerian conditions, while for Jupiter-like planets  estimates give that about $10^4$ orbital periods are needed to half the orbital size. Timescales of planet migration are strongly dependent on the relative position of the planet with respect to the shock region near the centrifugal barrier of the disc flow.
\end{abstract}

\begin{keywords}
 planetary systems: formation -- planetary systems: protoplanetary discs
\end{keywords}

\section{Introduction}
\label{Introduction}
Currently, nearly 300 extra-solar planets have been detected (http://www.exoplanets.eu/), and most of them are massive (Jupiter-like) planets with small (fractions of AU) semi-major axes (see \citet{Perryman00} and \citet{Udry07} for a review on detection methods and on the main properties of exo-planets). Theoretical models for the formation of planetary systems are then mostly based on the crucial role of an accretion disc around a forming young star, where the disc provides both the material for the forming planets and the interactions responsible for the migration of the protoplanets towards the central star.
The need for protoplanetary migration comes from the difficulty of an ``in situ'' formation in such close to the star positions, so that the detailed understanding of migration mechanisms appears mandatory in order to build self-consistent models for planetary systems formation.
Some analytic studies on the matter \citep{Goldreich79, Goldreich80, Lin84, Lin86a, Lin86b, Ward97} suggest that gravitational torques due to resonant non-axisymmetric structures could act to transfer angular momentum between the protoplanets and the disc (for a discussion of other alternative migration models see \citet{Delpopolo05} and references therein).
The gravitational disturbance exerted by the protoplanet on the disc produces spiral density waves which break the axial symmetry of the disc structure. Perturbation theory allows an estimate of the net torque that this perturbed disc exerts back onto the planet at ``Lindblad resonance sites'' \citep{Goldreich80, Ward97}.
This disc-planet interaction induces also a repulsion of material on either side of the protoplanet's orbit, with a possible formation of a density gap, depending on the interaction strength.
Two main migration types are then distinguished: {\it type I} \citep{Tanaka02}, when an Earth-like planet, not capable of opening a density gap in the disc is embedded in the disc structure. In this case the migration rate should be proportional to the protoplanet mass;
{\it type II} when a massive protoplanet, capable of opening and maintaining a density gap, remains trapped in this gap. In this case migration is due to the viscosity dominated evolution of the disc. 
In both cases it is possible that the lifetime of migration is shorter than the lifetimes of planet formation or disc dissolution, causing the collapse of the planet toward the star.
Accurate numerical simulations are necessary to better understand the matter \citep{Papaloizou06}. 
The main aim of this work is to study the role of initial and boundary conditions on planetary migration as a consequence of planet-disc dynamic interaction during the initial formation phases of stellar systems. Several papers devoted to this theme, according to various fluid-dynamics schemes \citep{Artymowicz04,Dangelo02,Dangelo03,Dangelo06,Kley00,Papaloizou06,Shafer04}, showed that this mechanism should be very effective for the protoplanet dragging toward the central protostar in characteristic time-scales of the order of $10^5$ years.
These studies assume that the accretion disc is basically Keplerian and pressure forces are often neglected among the protostar-disc interactions. In Keplerian models, the Keplerian distribution of disc gas velocity is usually assumed as a reasonable hypothesis \citep{Cresswell06,ValBorro06,Kley00,Shafer04,Terquem00}. However, as other authors have already done for a wider context during the accretion disc formation phase \citep{Sollins05,Ulrich76} or for small sub-Keplerian correction due to the contribution of pressure gradients \citep{Paardekooper04},
 we find useful to explore here the role of sub-Keplerian flows  as an open possibility for pre-main sequence accretion, as well as their role in the radial transport of solid bodies.
Therefore sub-Keplerian flows in the formation of protoplanetary accretion discs are here analysed as a parametric study (see section \ref{Initial_conditions} for details on exact initial conditions).\\
In this paper non viscous Cartesian 2D SPH sub-Keplerian disc models are given, with planetary migration times both for Jovian and for  Terrestrial planets. Models of Jovian or Terrestrial planets, interacting with an initial Keplerian accretion disc, are also produced for a strict comparison.\\
The coming out of shock fronts and complex structures, both steady and progressive in sub-Keplerian turbulent accretion flows in AGN, was described in \citet{Chakrabarti93}, \citet{Molteni94}, \citet{Lanzafame98}, as well as pulsed or periodic outflows in \citet{Lanzafame08} as a consequence of collisional interactions in the disc surrounding the central accreting object in axial symmetry.
In such environments (AGN or equivalently protostellar systems), the potential importance of the centrifugal barrier in developing outflow structures in turbulent collisional fluid-dynamics, was successfully  described in those papers. This implies that the possibility of a slower dragging toward the central star cannot be excluded in principle. Therefore, as a secondary aim, the purpose of the paper is the methodical search for those initial and boundary conditions determining results showing how Jovian and Terrestrial protoplanets could stay in orbits surrounding the central stars at distances of the order of that of Venus or Mercury.\par
In section \ref{model_boundary} we discuss the initial and boundary conditions characterising our models. In sec. \ref{results} we describe our results, and we discuss them in section \ref{discussion}.

\section{Models features and boundary conditions}
\label{model_boundary}

\subsection{Models}
\label{models}
The simulations presented in this work are based on a two-dimensional model which makes use of the Smoothed Particle Hydrodynamics (SPH) numerical method \citep{Monaghan92}. The choice of using the SPH method is funded on the grounds of its Lagrangian nature and its ability to easily tackle hydrodynamic problems with free surfaces. The ability of many different numerical schemes, both grid-based and Lagrangian, to correctly describe planet-gas interactions was tested by \citet{ValBorro06}, showing that SPH is not significantly less performant than other methods.
Given the lack of a wide range of case studies on the matter in SPH, we believe that our work could give an interesting contribution to improve the basic knowledge on disc-protoplanet interaction.\\
The code does not contain a specific radiation treatment, but the used equation of state: $P = (\gamma -1) \rho u$, where $u$ is the thermal energy per unit mass, includes the polytropic index $\gamma$, which can be adjusted to values lower than 5/3 if radiation, partial molecular dissociation or partial ionisation effects are present, giving the overall effect of a higher compressibility to the gas. Current calculations are performed with the simple choice of $\gamma=5/3$, but we intend to further investigate this matter in a future work.\\
The model is built in Cartesian coordinates with dimensionless quantities. Reference physical units are:
\begin{itemize} 
\item[-] the initial stellar mass $M_0$;
\item[-] the stellar radius $R_0$;
\item[-] the Keplerian period of an orbit of radius $R_0$ around a star of mass $M_0$,
$T_0 = \frac{2 \pi}{\sqrt{G M_0}} R_0^{3/2} $
\end{itemize}
With these choices, the following units can be used to transform our numerical results in physical quantities:
\begin{itemize}
\item[-] velocity: $v_0= \frac{1}{2 \pi} \sqrt{\frac{G M_0} {R_0}}$
\item[-] specific angular momentum: $j_0 = \frac{1}{2 \pi} \sqrt{G M_0 R_0}$
\item[-] energy per unit mass: $u_0= \frac{1}{4 \pi^2} \frac{G M_0}{R_0}$
\item[-] surface density: $\Sigma_0= \frac{M_0}{R_0^2}$
\end{itemize}
The initial conditions used to build the accretion disc are axially symmetric. The accretion disc is in fact created through the injection of particles at point-like positions (injectors) along circles concentric with the central star. Particles have initial tangential velocities set by choosing a value for the angular momentum per unit mass $j$. 
At some point, the accretion disc reaches a ``nearly'' steady state population of particles (time averaged since some variation of the number of particles and of the density distributions are actually observed during our simulations, due to the highly turbulent nature of our calculated flows) with $3\times10^5-5\times10^5$ particles, depending on the model, when using a smoothing length (spatial resolution parameter of the SPH method) $h=0.3$. At this stage, the planet is inserted and the model is evolved.
We included the following interactions in our models:
\begin{enumerate}
\item gravitational interaction between gas particles and central star;
\item gravitational interaction between planet and star;
\item gravitational interaction between particles and planet;
\item gas pressure between neighbouring gas particles;
\item artificial viscosity (pressure contribution) between gas particles;
\item gas particle capture by the planet (conservation of momentum is used to correct the planet speed before removing the particle) when gas particles approach the planet area (planet-particle distance less than $2h$).
\end{enumerate}

Additionally, in some of our models, we tried to add the above missing pressure interaction between the disc gas and the protoplanet. With the above specified conditions, our models do not have the necessary spatial resolution to treat the details of the interaction between the gas and the protoplanet surface. The simplest way to add pressure interaction in this case, is to add a further ``ad hoc'' SPH gas particle at the protoplanet position, giving to it the planet's velocity and the mass of the other surrounding gas particles. The density is then calculated according to the SPH method and the pressure force acting onto this particle is directly transferred to the planet itself. Though this approach, which is hereafter indicated as adding a ``pseudo-atmosphere'' to the protoplanet, may appear simplistic, it allows the introduction of the, otherwise totally missing, pressure interaction acting onto the planet.\\
No self-gravitation is included in the gas for computational reasons, and this is not believed to negatively affect our computations since the disc total mass is $10^{-11}\times$ the number of particles, which for $\sim 10^5$ particles means $10^{-6}$ (in units of the central star mass $M_0$), so that the dominating gravitational force is still that of the central star.
  Reported values for disc masses are in the range $10^{-3} - 10^{-1} ~ M_{\odot}$ on the grounds of observational constraints from T-Tauri stars \citep{Hartmann00, Terquem00}, but this value is usually distributed over a disc with a radius of $100-1000~AU$.
In order to compare this to the structure of our simulated disc, given the dimensionless nature of our simulation, we need to choose some physical reference units.
If  1 M$_\odot$ and 2-3 R$_\odot$ are taken for a typical T-Tauri star \citep{Bertout89}, the disc has a dimension of about $\sim 2$ AU. Therefore our simulations cover only the central part of a typical T-Tauri accretion disc, and the mass included in the simulation is in agreement with observational constraints.\par
Most accretion disc models in literature include viscosity, according to the disc model of \citet{Shakura73}. Even today, it is  hard to find a self-consistent way to give a correct value for the $\alpha_{SS}$ viscosity parameter.
We choose instead to not include physical viscous terms in the equations of our models, because we are here focused on the protoplanet migration mechanism based on gravitational and pressure forces and momentum transfer through gas particle captures, rather than to viscosity driven migration. 
Our models our non viscous, meaning that ar based on the Euler equations.
The absence of physical viscosity enhances the turbulent nature of our simulation, allowing us to correctly describe some effects due to turbulence like the development of non-axisymmetric structures in the disc (induced also by the presence of the protoplanet).
Turbulence might also produce stronger shock fronts when infalling gas particles hit the centrifugal barrier.\\
There are plans to include physical viscosity in a future work, in order to better explore the viscosity driven migrations (like in type II migration models) and viscosity contribution in general.

\subsection{Initial conditions}
\label{Initial_conditions}
Three choices are made here for the angular momentum per unit mass $j$: 18, 36, 54, for  particles injected at a radial distance of 130. All the three choices give a sub-Keplerian feature to the velocity distribution of the accretion disc (particles injected at a radial distance of 130 would have Keplerian velocity with $j \sim 70$), so that injected particles start decreasing their distance to the centre, until they ``hit'' the centrifugal barrier.
These choices are considered in the the framework of a parametric study, to test the effect of a wide range of possible conditions, from ``extreme'' ($j=18$) to ``weak'' ($j=54$) sub-Keplerian.
The mass of the particles is set to $10^{-11}$ (in units of $M_0$), which means that the total mass of the disc can be about $10^{-6} - 10^{-5}$. Calculations are performed with two masses for the planet: $10^{-7}$ (Earth-like planet) and $10^{-3}$ (Jupiter-like planet).
The computational domain is circular, extended in the xy plane with a radius of 150 (the star is located in the ``origin''), and the planet is located inside the disc plane, with the initial position at $X_p=100$ and $Y_p=0$, and an initial zero eccentricity.\\
\begin{figure}
\resizebox{\hsize}{!}{\rotatebox{-90}{\includegraphics[clip=true]{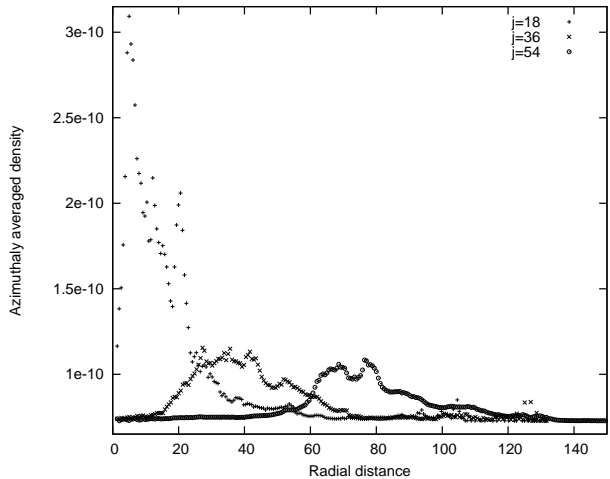}}}
 \caption{Azimuthally averaged density profiles of the sub-Keplerian accretion discs, just before planet insertion,  for $j=18,36,54$.}
\label{initial_density_profiles}
\end{figure}
The azimuthally averaged density profiles of the sub-Keplerian accretion discs, just before planet insertion, are plotted in figure \ref{initial_density_profiles} for the three cases ($j=18,36,54$). A Density peak is present near the centrifugal barrier in each case. For $j=18$ the peak is close to the star, while for higher $j$ values the density close to the star has values similar to those found at the disc outer edges. Density in SPH is computed according to local particle mass distribution. At the disc outer edge, particles are produced and injected in the stellar gravitational potential well. Instead particles are in a free fall very close to the central star, where gravitational potential rules the whole gas dynamics, and they are ``removed'' from the simulation (the mass of the particle is added to the star mass) when they reach a radial position equal to the star radius.\\
As far as the disc material accretion rate onto the central star, if 1 M$_\odot$ and 2-3 R$_\odot$ are taken for a typical T-Tauri star \citep{Bertout89}, we obtain for the three cases values of $2\times 10^{-6},~ 6\times 10^{-7},~ 3\times10^{-7}~ M_{\odot}/year$ with oscillations of about $30\%,80\%,100\%$. These values are in agreement with estimated accretion rates for T-Tauri stars \citep{Gullbring98, Hartigan95}.\\
In order to compare some of our results with more traditional models, some computations (two with an Earth-like planet and two with a Jupiter-like planet) are built using an initial exactly Keplerian distribution of gas particle velocities, with an initial uniform density over the entire disc area. For simplicity, we will refer to these models as ``Keplerian'', even if the velocity distribution of the gas particles cannot remain exactly Keplerian during the simulation, due to the pressure forces within the gas and to the interactions between the planet and the accretion disc.
These models will be used to test the eventual formation of a gap near the protoplanet location, and to compare some classical views on the subject \citep{Ward97} with our results obtained with an inviscid model.\\
A summary of the distinguishing features of our models is given in table \ref{models_table} together with identifying labels.
\begin{table}
 \begin{tabular}{|c | c c c |}
Label & $j$ & $M_p$ & pseudo-atmosphere \\
\hline
18a     & 18.0 &  $10^{-3}$ & no \\
18b     & 18.0 &  $10^{-3}$ & yes \\
18c     & 18.0 &  $10^{-7}$ & no \\
18d     & 18.0 &  $10^{-7}$ & yes \\
36a     & 36.0 &  $10^{-3}$ & no \\
36b     & 36.0 &  $10^{-3}$ & yes \\
36c     & 36.0 &  $10^{-7}$ & no \\
36d     & 36.0 &  $10^{-7}$ & yes \\
54a     & 54.0 &  $10^{-3}$ & no \\
54b     & 54.0 &  $10^{-3}$ & yes \\
54c     & 54.0 &  $10^{-7}$ & no \\
54d     & 54.0 &  $10^{-7}$ & yes \\
kepl\_a   & 71.6 &  $10^{-3}$ & no \\
kepl\_b   & 71.6 &  $10^{-3}$ & yes \\
kepl\_c   & 71.6 &  $10^{-7}$ & no \\
kepl\_d   & 71.6 &  $10^{-7}$ & yes \\
\hline  
 \end{tabular}
\caption{List of computed models, with their identifying labels (first column). The second column gives the specific angular momentum of injected gas particles, the third column gives the initial mass of the protoplanet and the fourth column indicates the eventual presence of the pseudo-atmosphere on the protoplanet.}
\label{models_table}
\end{table}
In the table, $M_p$ is the mass of the planet in units of the central star mass, and ``pseudo-atmosphere'' indicates the eventual presence of a gas SPH particle associated with the planet.\\

\section{Results}
\label{results}
\begin{figure}
\resizebox{\hsize}{!}{\includegraphics[clip=true]{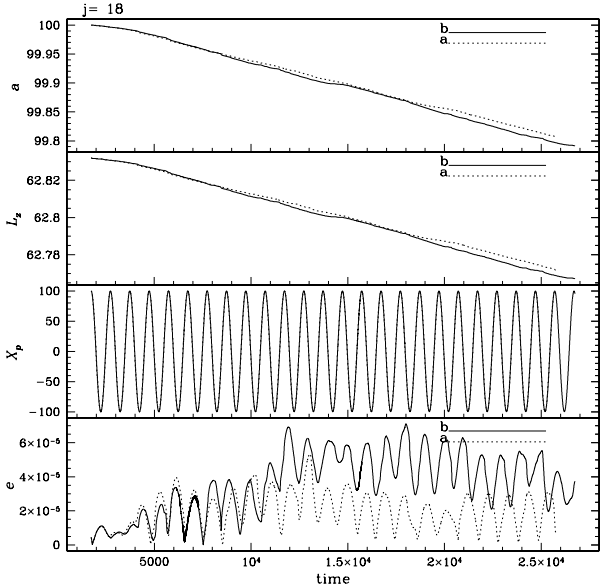}}
 \caption{Orbital parameters evolution for models 18a and 18b. The semi-major axis $a$, the specific angular momentum $l_z$ a, the $x$ coordinate of the protoplanet and the eccentricity $e$ are reported, as a function of time. The dotted and continuum graphs are for the two reported data sets, with and without the protoplanet pseudo-atmosphere.}
\label{l18_a_b}
\end{figure}

\begin{figure}
\resizebox{\hsize}{!}{\includegraphics[clip=true]{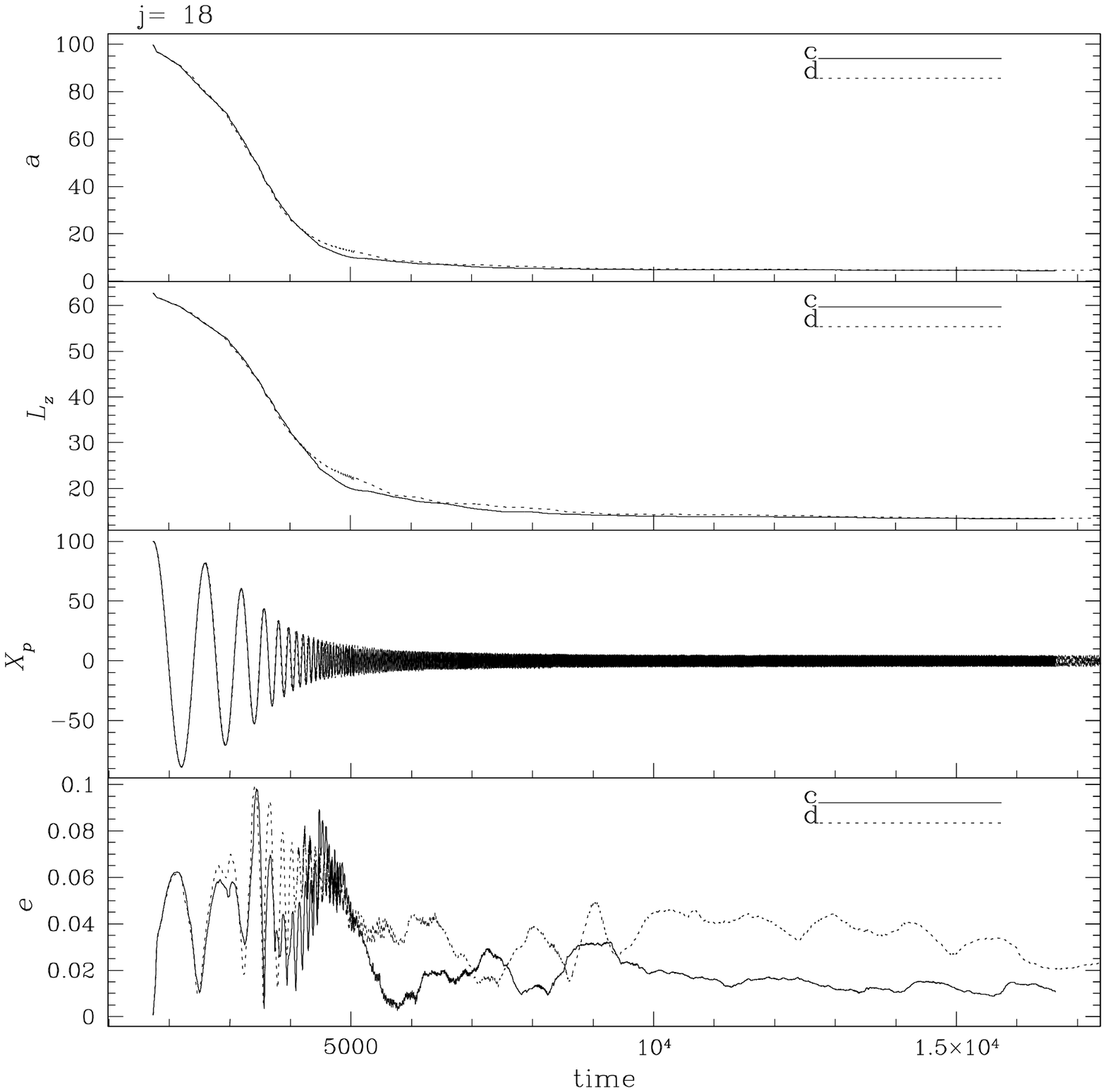}}
 \caption{Orbital parameters evolution for models 18c and 18d. See fig. \ref{l18_a_b} caption for description.}
\label{l18_c_d}
\end{figure}

\begin{figure}
\resizebox{\hsize}{!}{\includegraphics[clip=true]{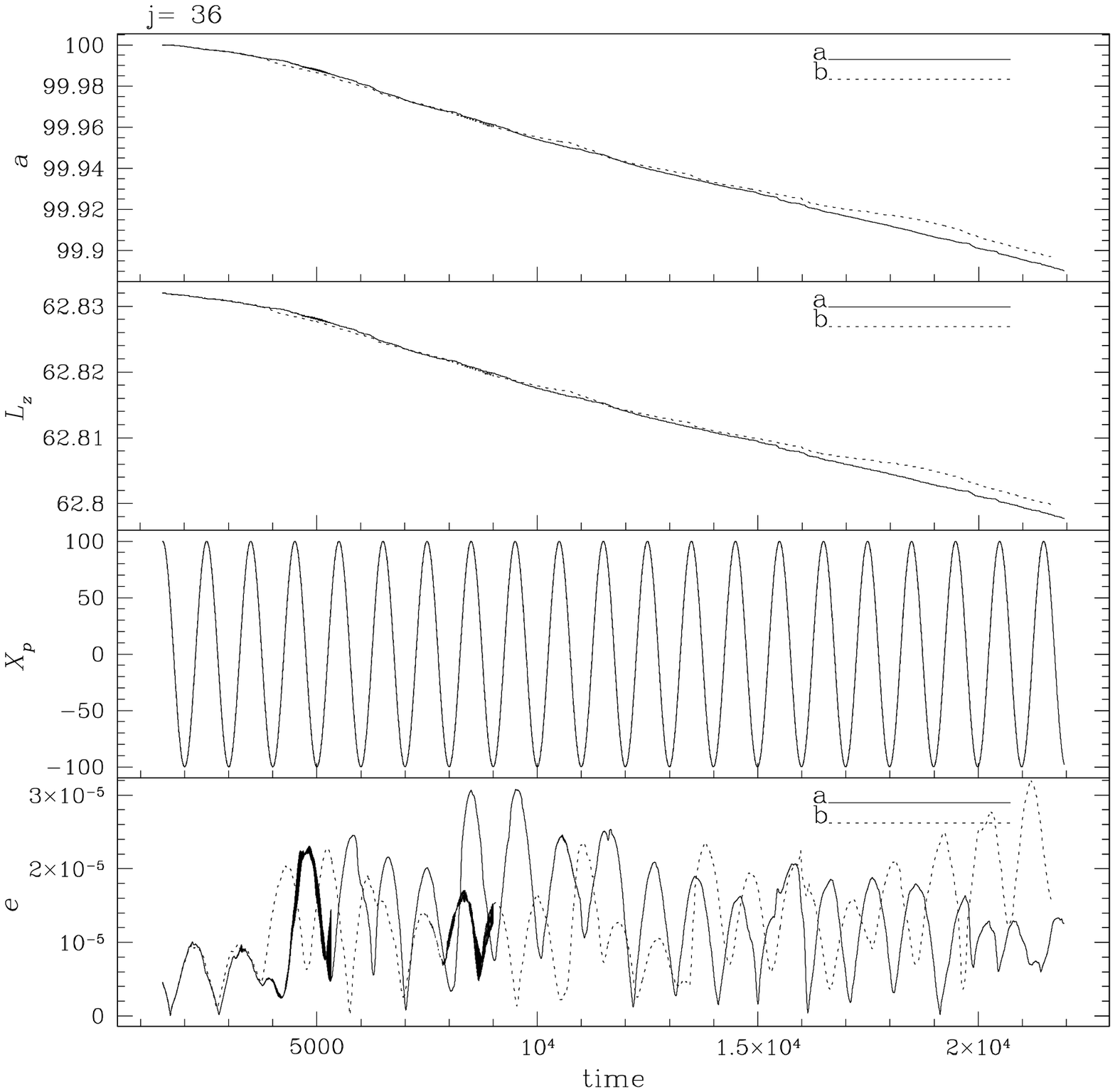}}
 \caption{Orbital parameters evolution for models 36a and 36b. See fig. \ref{l18_a_b} caption for description.}
\label{l36_a_b}
\end{figure}

\begin{figure}
\resizebox{\hsize}{!}{\includegraphics[clip=true]{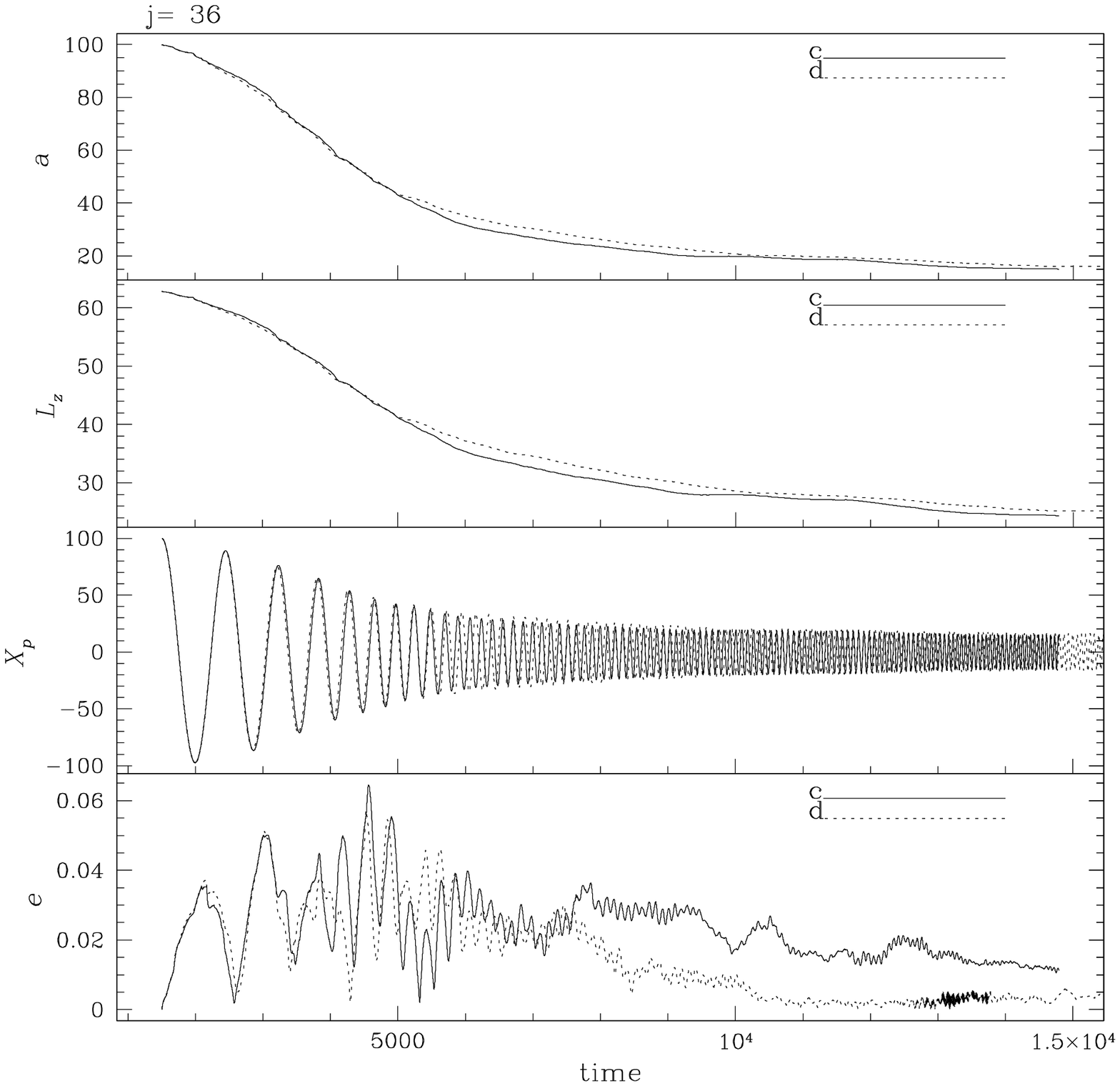}}
 \caption{Orbital parameters evolution for models 36c and 36d. See fig. \ref{l18_a_b} caption for description.}
\label{l36_c_d}
\end{figure}

\begin{figure}
\resizebox{\hsize}{!}{\includegraphics[clip=true]{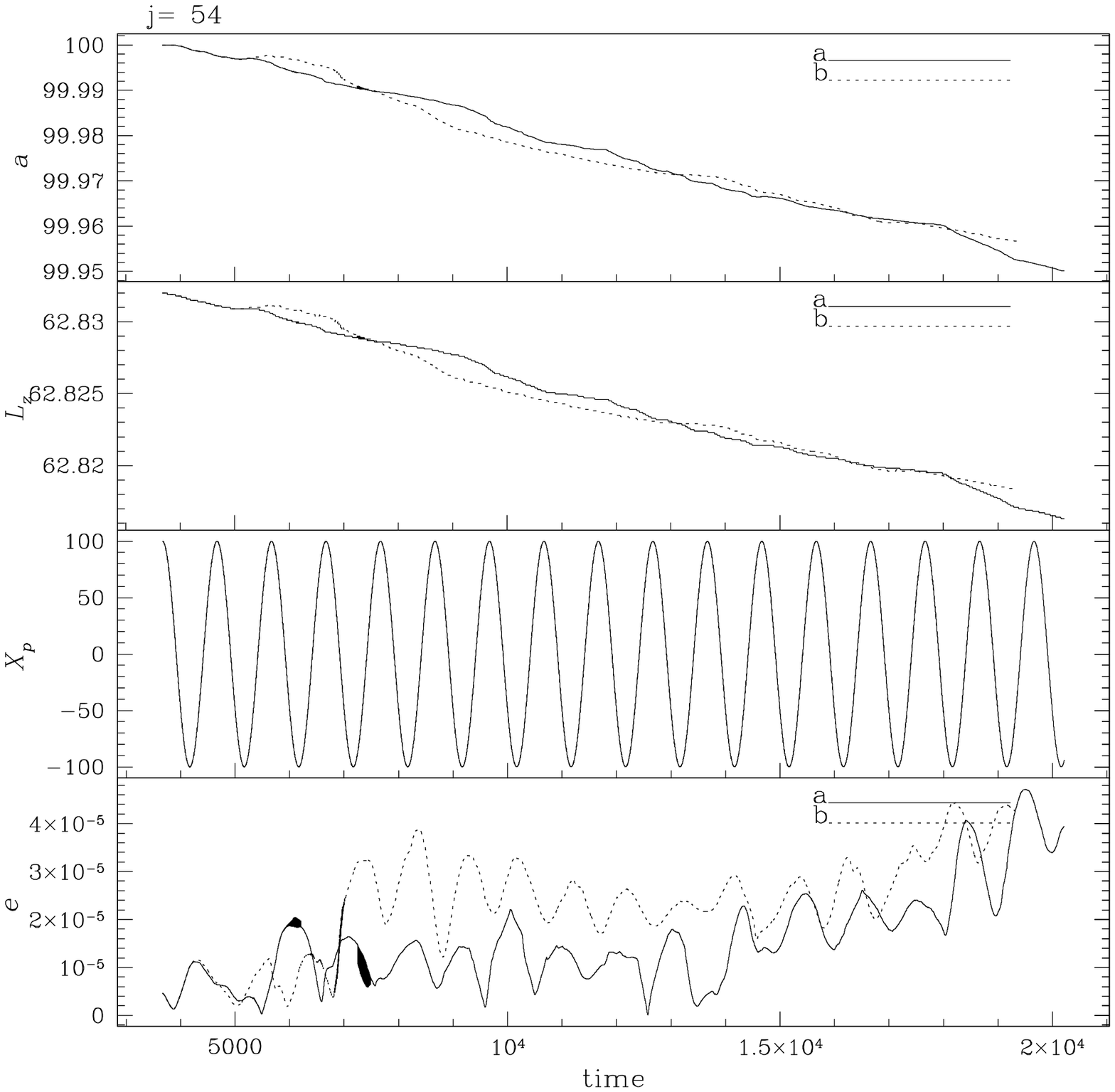}}
 \caption{Orbital parameters evolution for models 54a and 54b. See fig. \ref{l18_a_b} caption for description.}
\label{l54_a_b}
\end{figure}

\begin{figure}
\resizebox{\hsize}{!}{\includegraphics[clip=true]{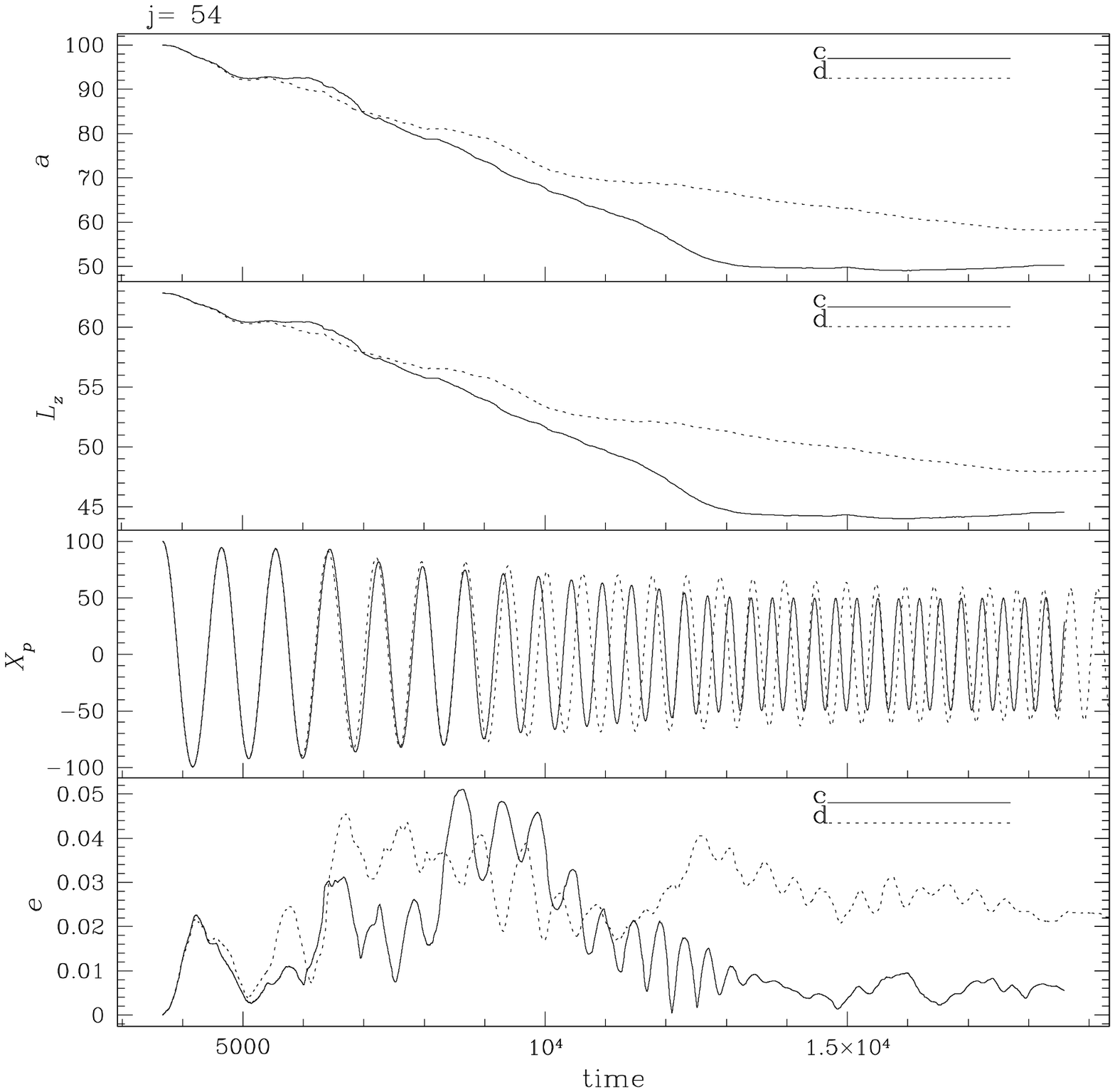}}
 \caption{Orbital parameters evolution for models 54c and 54d. See fig. \ref{l18_a_b} caption for description.}
\label{l54_c_d}
\end{figure}

\begin{figure}
\resizebox{\hsize}{!}{\includegraphics[clip=true]{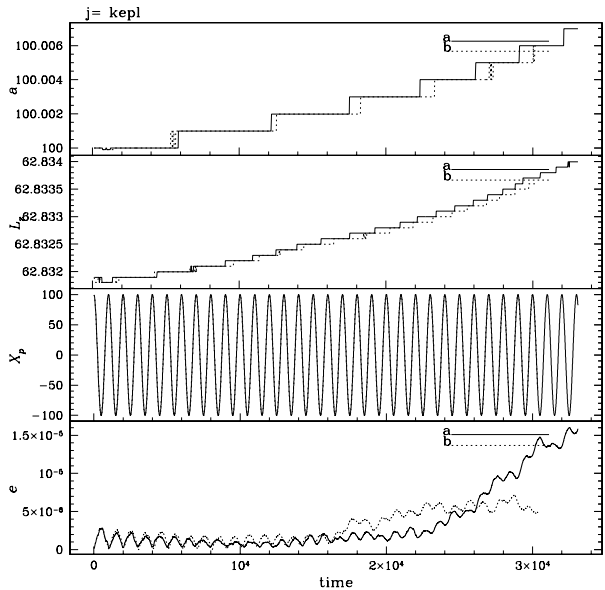}}
 \caption{Orbital parameters evolution for the Keplerian models and the Jovian planet. See fig. \ref{l18_a_b} caption for description.}
\label{Keplerian_a_b}
\end{figure}

\begin{figure}
\resizebox{\hsize}{!}{\includegraphics[clip=true]{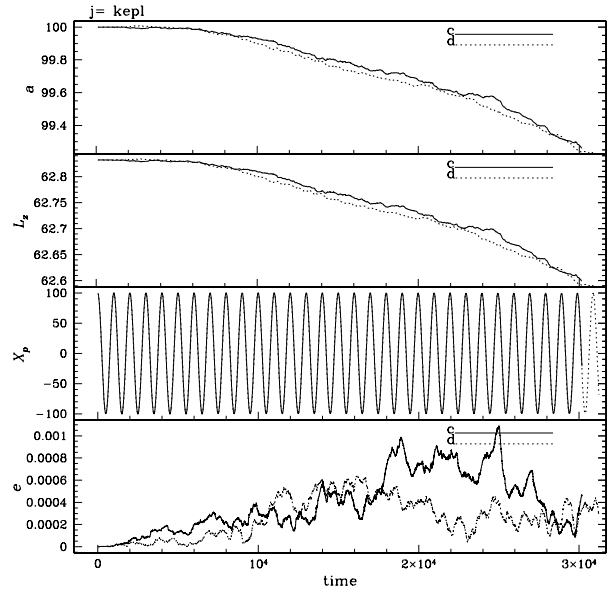}}
 \caption{Orbital parameters evolution for the Keplerian models and the Earth-like planet. See fig. \ref{l18_a_b} caption for description.}
\label{Keplerian_c_d}
\end{figure}

The turbulent nature of our disc models is evidenced by the high value of the Reynolds number, evaluated by using the artificial viscosity coefficient $\nu_{num}=c_s h$ \citep{Molteni91}, especially at the disc outer edge, where we have values of about 1200, 2400, 3600 for $j=18,36,54$ respectively. Instead, in the innermost regions, where the disc gas temperature is higher, we still evaluate Re values of about $300 \div 1000$.\\
Figures \ref{l18_a_b}-\ref{Keplerian_c_d} show the evolution of the orbital parameters for all models. The semi-major axis $a$ as a function of time is shown in the top rectangle of each figure; the angular momentum of the protoplanet per unit mass $l_z$, the coordinate $x$ and the eccentricity $e$ are shown in the following rectangles, respectively. The coordinate $x$ is plotted to show the orbital phase, which allow us to count protoplanet orbits.
Each figure includes the results from two models, with (models with labels ``b'' and ``d'') and without (models ``a'' and ``c'') the pseudo-atmosphere gas particle for the planet.
The main feature that emerges from the sub-Keplerian models is a downward migration of the protoplanet, the speed of which being strongly dependent on the planet mass. Earth-like planets migrate in a few orbits, especially in the $j=18$ case, while for the Jupiter-like planets, an extrapolation of our results (which are limited to 20-30 orbits), suggests that a few thousands orbits are necessary to move the planets towards the central star.\\
Of course, the above suggested extrapolation would require a linear behaviour of the semi-major axis (or the specific angular momentum of the planet) with respect to time. But the results obtained for the less massive planet suggest that the migration rate asymptotically decreases for all the three values of the sub-Keplerian angular momentum of the injected particles.
All the sub-Keplerian models with the Jovian planet lack the ``classical'' gap opening around the planet. This is not believed to be a problem as the type II migration mechanism, with the planet blocked inside the induced density gap, was developed in the framework of a Keplerian gas distribution in the accretion disc and within a viscosity based fluid model. Therefore the migration mechanism suggested within our model cannot be directly associated to the classical type I - type II distinction.\\
For the Keplerian models, fig. \ref{Keplerian_a_b} shows that the Jupiter-like planet migration slowly proceeds outwards, despite the development of density gap soon after the introduction of the planet in the initially totally Keplerian gas distribution (fig. \ref{Keplerian_disc_Jovian}), while the Earth-like planet (fig. \ref{Keplerian_c_d}) migrates downward, with a slower rate when compared to that of sub-Keplerian discs. We have in fact $0.6\%$ variation in $a$ in about $30$ orbits, while, with the same number of orbits, the sub-Keplerian discs show a $\sim 90\%; ~ 80\%;~ 50\%$ decrease of the planet semi-major axis, for $j_z=~18;~36;~54$ respectively.\\
As far as the influence of the pseudo-atmosphere is concerned, the difference between the results obtained with and without this model structure is of the order of $\sim 10^{-2}\%$ with the Jovian protoplanet and of $\sim 1-4\%$ with the Terrestrial protoplanet.
In most cases we have an initial faster migration rate with the pseudo-atmosphere, but the situation is inverted after a few protoplanet's orbits. This is clearer in models with a Terrestrial protoplanet for which migration rate is much faster in all models.
A significant difference is registered between the $54c$ and $54d$ models, where a $\sim 10-15 \%$ difference is obtained in the semi-major axis $a$ in advanced migration phases. The model without the pseudo-atmosphere ($54c$) suffers the fastest migration rate at some point.\\

\begin{figure}
\resizebox{\hsize}{!}{\rotatebox{-90}{\includegraphics[clip=true]{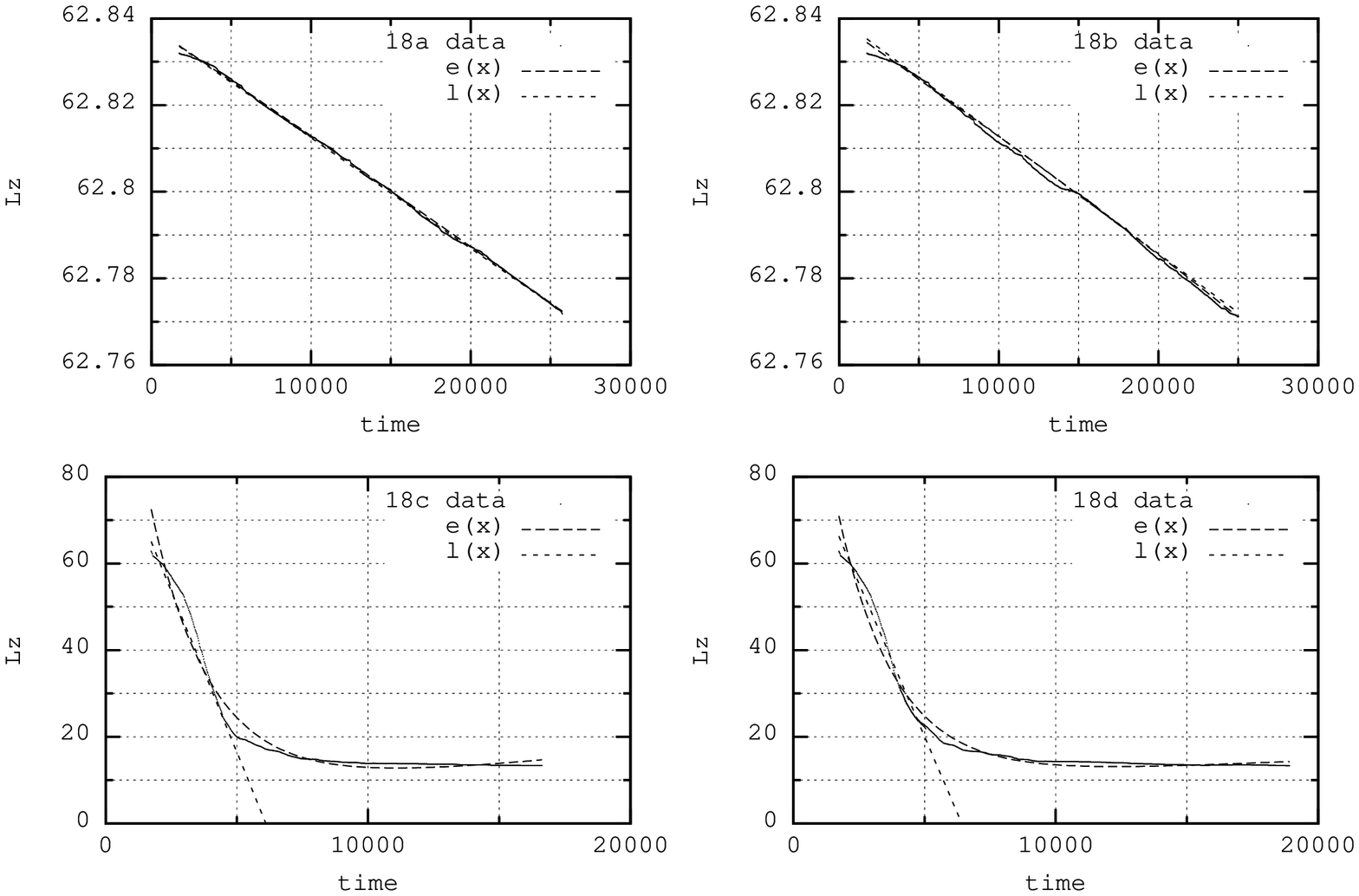}}}
 \caption{Data and exponential best-fit for angular momentum in $j=18$ models}
\label{l18_lz}
\end{figure}

\begin{figure}
\resizebox{\hsize}{!}{\rotatebox{-90}{\includegraphics[clip=true]{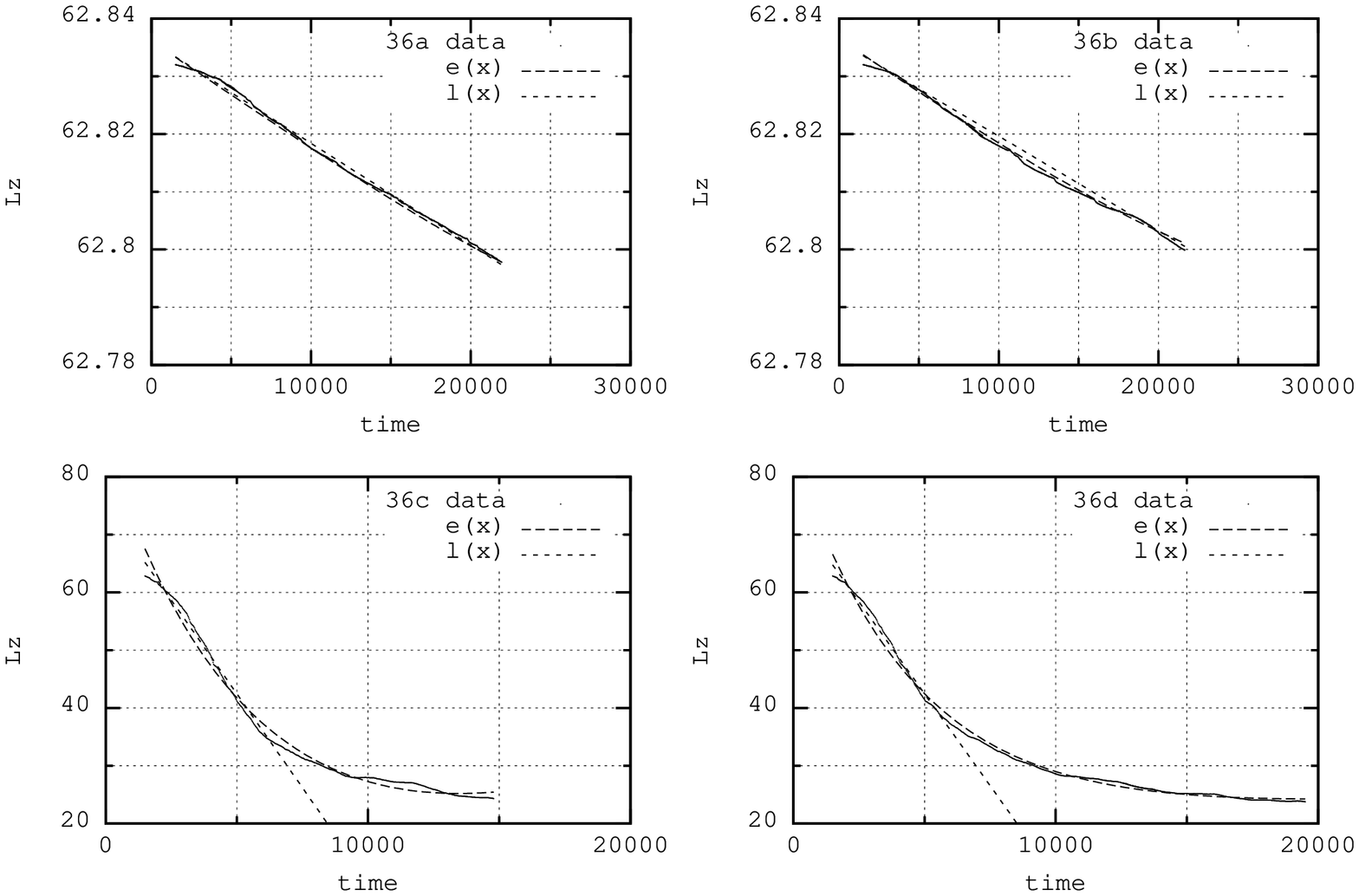}}}
 \caption{Data and exponential best-fit for angular momentum in $j=36$ models}
\label{l36_lz}
\end{figure}

\begin{figure}
\resizebox{\hsize}{!}{\rotatebox{-90}{\includegraphics[clip=true]{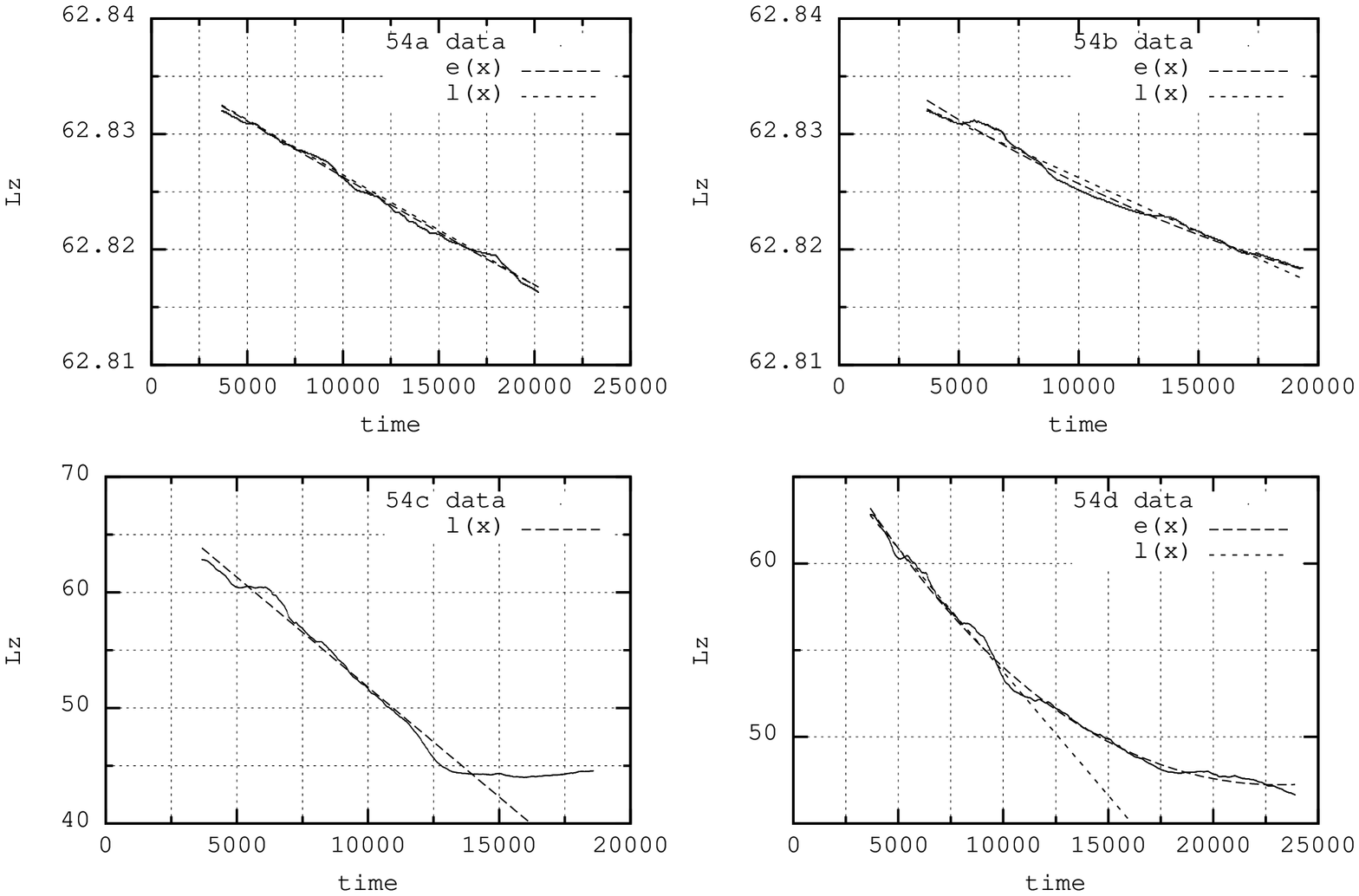}}}
 \caption{Data and exponential best-fit for angular momentum in $j=54$ models}
\label{l54_lz}
\end{figure}

\begin{figure}
\resizebox{\hsize}{!}{\rotatebox{-90}{\includegraphics[clip=true]{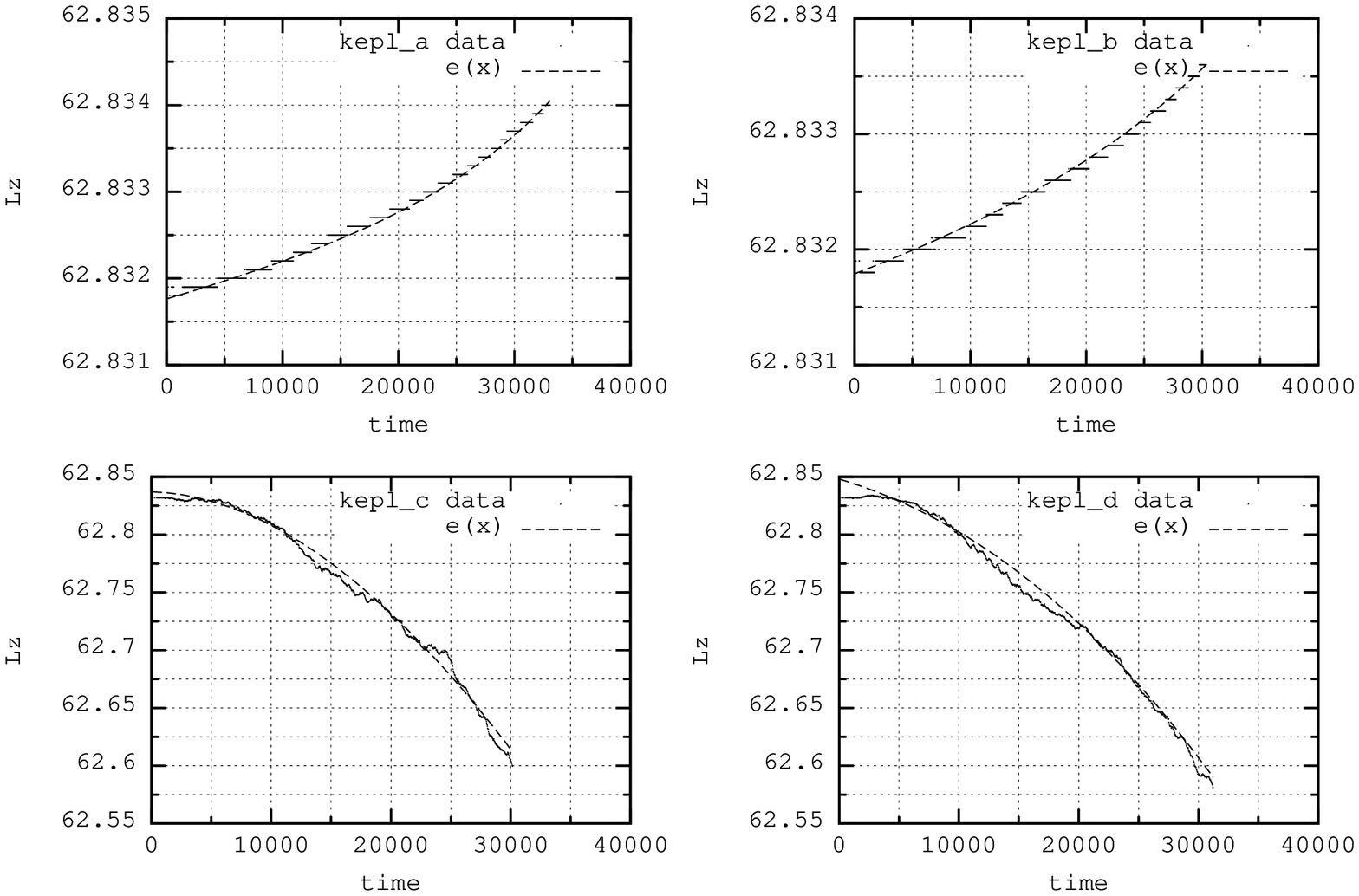}}}
 \caption{Data and exponential best-fit for angular momentum in Keplerian models}
\label{Keplerian_lz}
\end{figure}

In order to draw a semi-analytic behaviour in the orbital parameters evolution of the planets in our models, best-fits have been computed for the specific angular momentum $j$ (figures \ref{l18_lz}-\ref{Keplerian_lz}) to have an estimate of the time derivatives and decrease lifetime of the angular momentum.
Two kind of fits have been used: linear and exponential, according to the following schemes:

\begin{equation}
j_z = a*exp(b\cdot t)+c*exp(d\cdot t) + e
\label{exp_template}
\end{equation}

\begin{equation} 
j_z = a t + b
\label{linear_template}
\end{equation}
Tables \ref{exp_fit} and \ref{linear_fit} report the best-fit parameters obtained for the various models.

\begin{table*}
\begin{tabular}{ c c c c c c c c }
 model & a & b & c & d & e & $1/b$ (years) & $1/d$ (years)\\
\hline
18a & $62.87$ & $-3.4\times 10^{-8}$ & -$3.412\times 10^{-2}$ & $1.059\times 10^{-5}$ & $2.2\times 10^{-2}$ &
$-2.66\times10^4$ & $8.54\times 10$\\
18b & $-1.054\times 10^{-3}$ & $7.134\times 10^{-5}$ & $62.84$ & $-3.994\times 10^{-8}$ & $0.0$ &
$1.27\times 10$ & $-2.27\times 10^4$\\
18c & $138.4$ & $-4.426\times 10^{-4}$ & $7.672$ & $3.873\times 10^{-5}$ & $0.0$ &
$-2.04$ & $2.34\times 10$\\
18d & $133.1$ & $-4.547\times 10^{-4}$ & $10.12$ & $1.822\times 10^{-5}$ & $0.0$ &
$-1.99$ & $4.97\times 10$\\
36a & $62.84$ & $-3.02\times 10^{-8}$ & $1.228\times 10^{-4}$ & $1.519\times 10^{-4}$ & $-4.0\times 10^{-3}$ &
$-3.00\times 10^4$ & $5.96$\\
36b & $62.83$ & $-3.209\times 10^{-8}$ & $1.538\times 10^{-3}$ & $8.522\times 10^{-5}$ & $5.0\times 10^{-3}$ &
$-2.82\times 10^4$ & $1.06\times 10$\\
36c & $77.65$ & $-1.842\times 10^{-4}$ & $7.86$ & $6.437\times 10^{-5}$ & $0.0$ &
$-4.91$ & $1.41\times 10$\\
36d & $63.92$ & $-2.191\times 10^{-4}$ & $20.37$ & $7.021\times 10^{-6}$ & $0.0$ &
$-4.13$ & $1.29\times 10^2$\\
54a & $2.84$ & $-1.557\times 10^{-8}$ & $7.47\times 10^{-6}$ & $2.087\times 10^{-4}$ & $-4.0\times 10^{-3}$ &
$-5.81\times 10^4$ & $4.33$\\
54b & $0.1758$ & $-1.734\times 10^{-5}$ & $62.66$ & $2.504\times 10^{-8}$ & $2.2\times 10^{-3}$ &
$-5.22\times 10$ & $3.61\times 10^4$\\
54c & - & - & - & - & - & - & -\\
54d & $56.29$ & $-4.93\times 10^{-5}$ & $14.5$ & $3.032\times 10^{-5}$ & $0.0$ &
$-1.84\times 10$ & $2.98\times 10$\\
kepl\_a & $62.83$ & $5.484\times 10^{-10}$ & $6.264\times 10^{-5}$ & $8.953\times 10^{-5}$ & $1.7\times 10^{-3}$ &
$1.65\times 10^6$ & $1.01\times 10$\\
kepl\_b & $62.83$ & $4.825\times 10^{-10}$ & $1.375\times 10^{-4}$ & $6.635\times 10^{-5}$ & $1.65\times 10^{-3}$ &
$1.87\times 10^6$ & $1.36\times 10$\\
kepl\_c & $65.74$ & $-6.256\times 10^{-7}$ & $-2.903$ & $-1.407\times 10^{-5}$ & $0.0$ &
$-1.45\times 10^3$ & $-6.43\times 10$\\
kepl\_d & $-1.982$ & $1.227\times 10^{-5}$ & $64.83$ & $3.28\times 10^{-7}$ & $0.0$ &
$7.37\times 10$ & $2.76\times 10^3$\\
\hline
\end{tabular}
\caption{Exponential best-fit parameters for the specific angular momentum time evolution. A rough estimate of migration times is given by the inverse of the two parameters $b$ and $d$, dimensioned values in {\it years} are given in 7th and 8th columns, calculated under the hypothesis of 1 $M_\odot$ star with a radius of $\sim2~R_\odot$.}
\label{exp_fit}
\end{table*}

\begin{table}
\begin{tabular}{ c c c c c }
 model & a & b & $\tau$ (adim) & $\tau$ (years)\\
\hline
18a & $-2.555\times 10^{-6}$ & 62.838 & $2.46\times 10^7$ & $2.23\times 10^4$\\
18b & $ -2.717\times 10^{-6}$ & 62.84 & $2.31\times 10^7$ & $2.09\times 10^4$\\
18c & $-1.48685\times 10^{-2}$ & 93.2653 & $4.23\times 10^3$ & 3.83\\
18d & $-141492\times 10^{-2}$ & 90.8171 & $4.44\times 10^3$ & 4.02\\
36a & $-1.762\times 10^{-6}$ & 62.836 & $3.57\times 10^7$ & $3.23\times 10^4$\\
36b & $-1.638\times 10^{-6}$ & 62.836 & $3.84\times 10^7$ & $3.47\times 10^4$\\
36c & $-6.49898\times 10^{-3}$ & 74.9447 & $9.67\times 10^3$ & 8.75\\
36d & $ -6.37015\times 10^{-3}$ & 74.307 & $9.86\times 10^3$ & 8.92\\
54a & $-9.533\times 10^{-7}$ & 62.836 & $6.59\times 10^7$ & $5.96\times 10^4$\\
54b & $-9.36912\times 10^{-7}$ & 62.8356 & $6.71\times 10^7$ & $6.07\times 10^4$\\
54c & $-1.90167\times 10^{-3}$ & 70.8222 & $3.30\times 10^4$ & 29.9\\
54d & $-1.42529\times 10^{-3}$ & 68.0032 & $4.41\times 10^4$ & 39.9\\
\hline
\end{tabular}
\caption{Linear best-fit parameters for the specific angular momentum time evolution. For some cases, the linear fit is calculated over a limited time interval, during which the angular momentum evolution is linear. The linear fits are not present for the Keplerian models since there is no time interval during which the evolution of the angular momentum is linear. Migration times estimated with the formula $l_z/\dot{l_z}$ are reported in the last two columns: in the fourth column the dimensionless time is reported, while in the last column the same time is converted in {\it years} by choosing 1 $M_\odot$ and 2 $R_\odot$ as physical reference units.}
\label{linear_fit}
\end{table}

In table \ref{exp_fit} the last two columns give the inverse values of the parameters $b$ and $d$, and they have been dimensioned in years by choosing 1 $M_\odot$ and $2~R_\odot$ as reference units. Of the two time-scales estimated in this way, the most important is the one with the corresponding highest ``multiplying factor for the exponential'' ($a$ or $c$). For Jupiter-like planets these estimates must be taken with some care, as the simulations did not go enough far in time to have a clear path of the migration process, so that migration timeline of these models could be different than what expected from extrapolations.\\
In table \ref{linear_fit} the 4th column gives a dimensionless migration time, evaluated as $l_z/\dot{l_z}$, and the last column gives the same time converted in years by choosing the same reference units as in table \ref{exp_fit}.\\
For models $a$ and $b$ (Jovian planet) a linear fit already gives a good description of the angular momentum evolution of the planet, but an exponential fit has also been performed because, on a longer time-scale (a few thousands orbits), the curve showing  $j$ as a function of time could resemble that of the $c$ and $d$ models. We are therefore able to analyse the different migration lifetimes through comparable parameters (exponential lifetime or linear costant derivative).\\
For some of $c$ and $d$ models, the behaviour is initially linear and it becomes more exponential once the protoplanet reaches a position closer to the star. In these cases, the linear fit has been done for a limited time interval during which the evolution does not show a significant migration rate variation.
A peculiar case is the $54c$ model where an initial linear behaviour with negative derivative is followed by a nearly complete migration stop. The exponential fit has not been evaluated for this model. \\

\begin{figure}
\resizebox{\hsize}{!}{\includegraphics[clip=true]{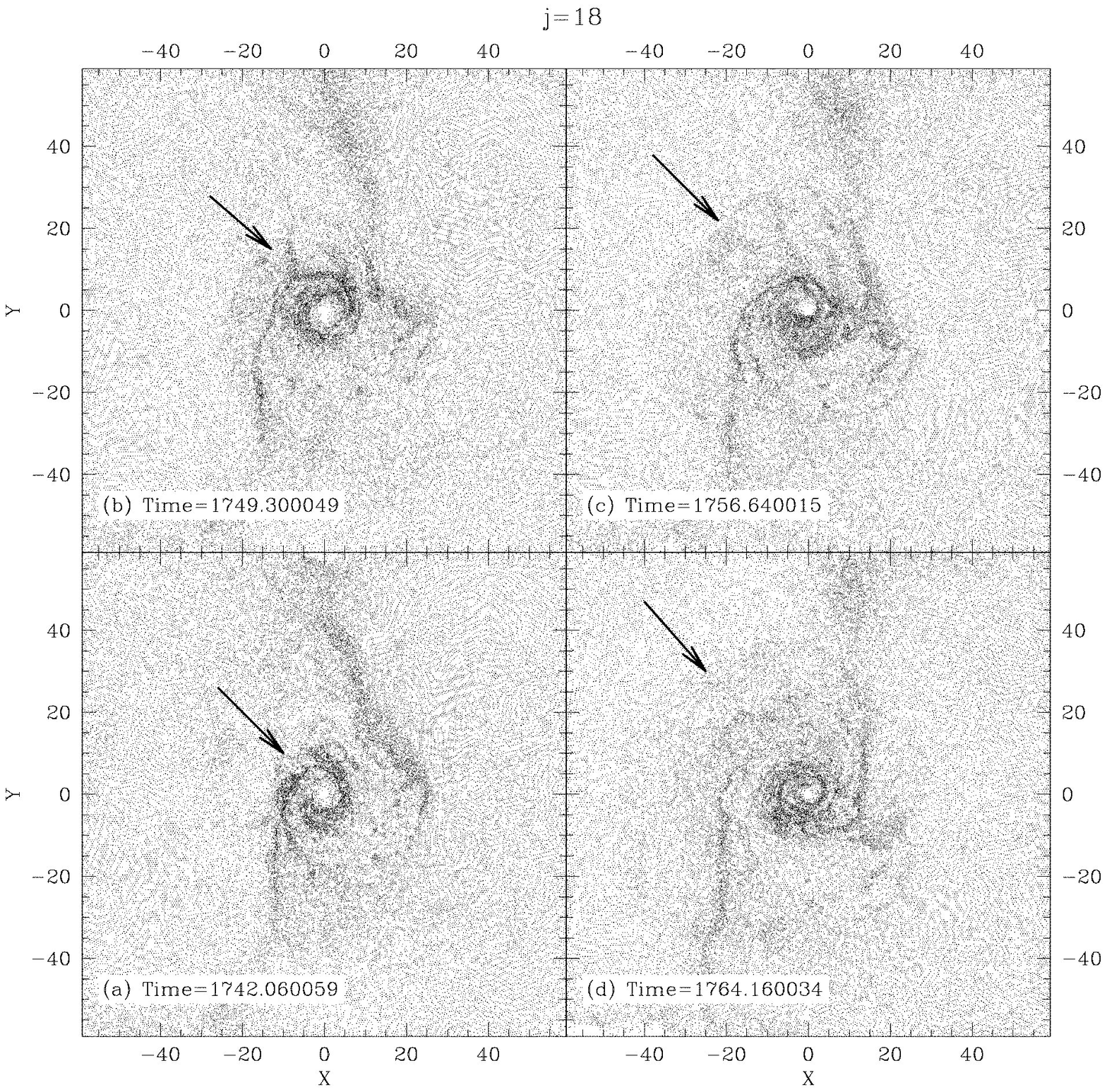}}
 \caption{Wave propagation in the $j=18$ accretion disc. The four snapshots are ordered according to increasing time, starting from the bottom-left square in clock-wise order. The arrows indicate the position of the shock front.}
\label{wave_l18}
\end{figure}

\begin{figure}
\resizebox{\hsize}{!}{\includegraphics[clip=true]{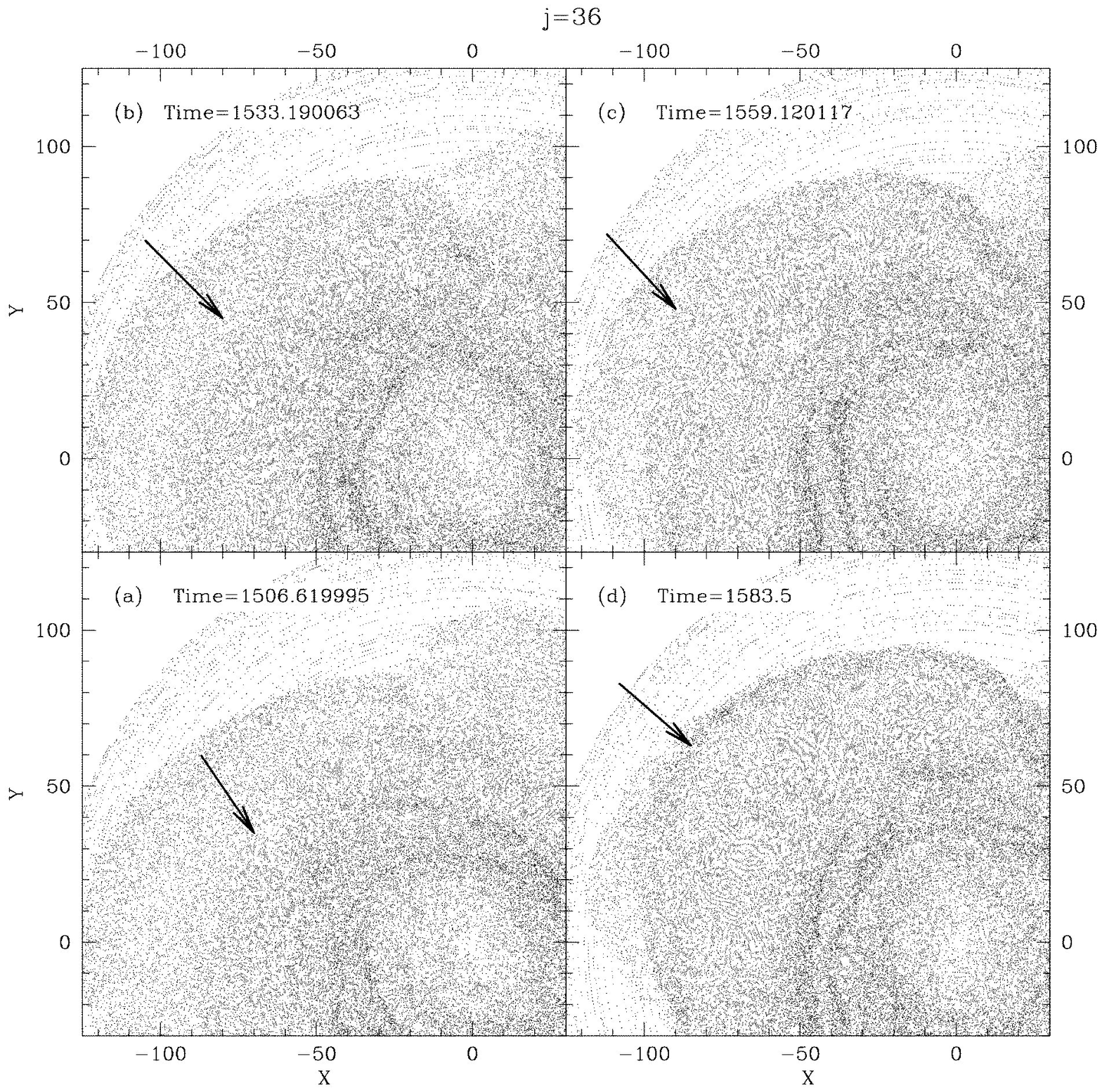}}
 \caption{Wave propagation in the $j=36$ accretion disc. The four snapshots are ordered according to increasing time, starting from the bottom-left square in clock-wise order. The arrows indicate the position of the shock front.}
\label{wave_l36}
\end{figure}

\begin{figure}
\resizebox{\hsize}{!}{\rotatebox{-90}{\includegraphics[clip=true]{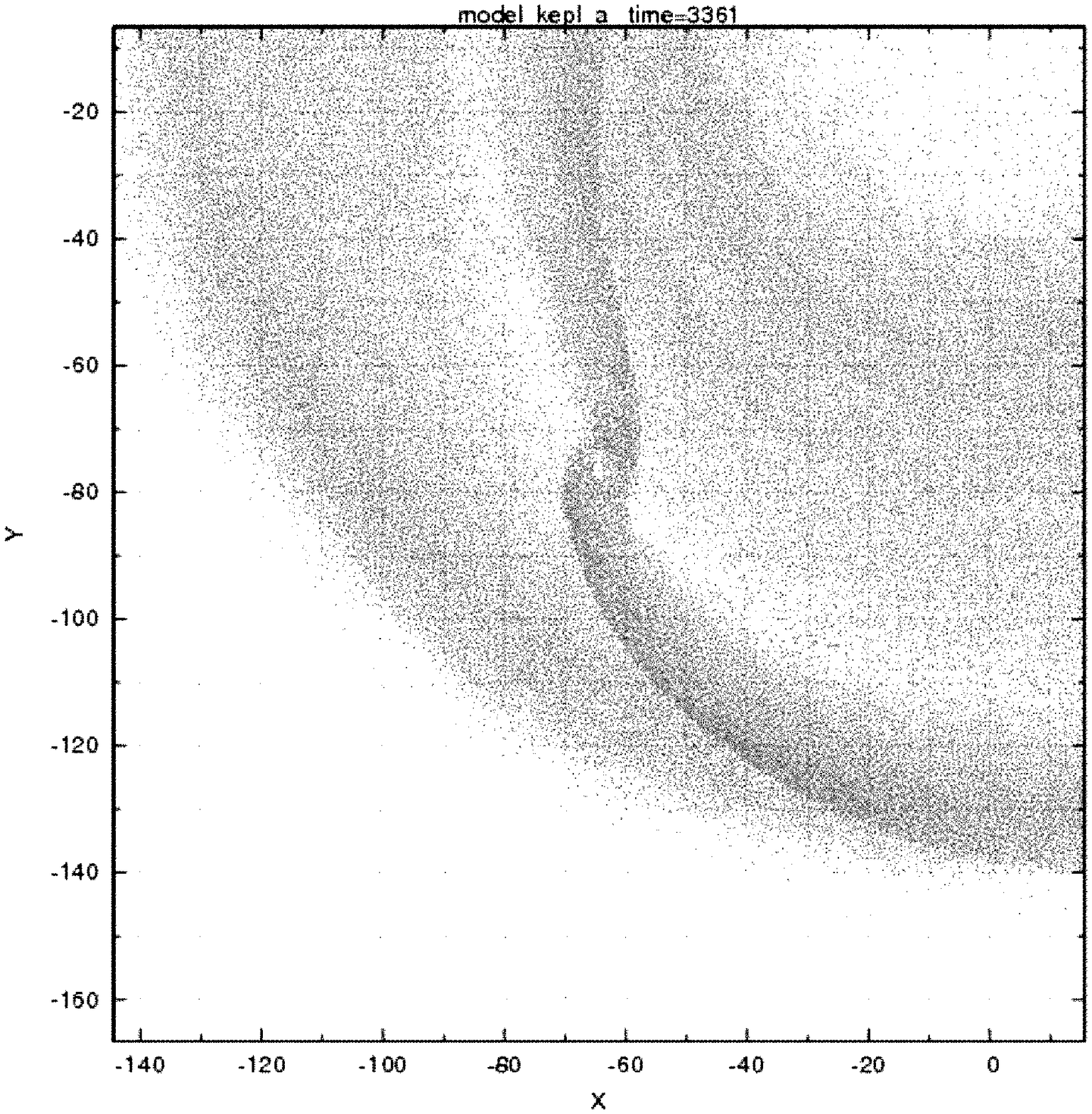}}}
 \caption{Distribution of the gas particles around a Jupiter-like protoplanet in the initially Keplerian accretion disc. This snapshot corresponds to a $\Delta t\sim3300$ after the artificial formation of a purely Keplerian velocity distribution, which corresponds to a bit more than 3 orbits of the protoplanet.}
\label{Keplerian_disc_Jovian}
\end{figure}

Propagation of waves coming from the central zone of the disc has been observed, at least for the $j=18$ and $j=36$ models. Some snapshots showing this phenomenon are plotted in figures \ref{wave_l18} and \ref{wave_l36}.
The same is not clearly observed in the other models, the cause being probably the high angular momentum of particles.
A strong sub-Keplerian condition for injected particles makes them rapidly fall towards the centrifugal barrier, and the subsequent bouncing-back could be at the origin of these waves.
Details on this propagation of shock waves in systems with a central accreting object was described by \citet{Lanzafame08} and references therein. A proper combination of sub-Keplerian gas flow and viscosity parameter values can create a complex shock structure with nearly periodical radial motions of the shock front. Periods of quiet piling up of particles near and below the centrifugal barrier are followed by the development of a shock front propagating outwards. In low or no viscosity regimes, the position of the shock front is generally closer to the central accreting object.

Together with the above specified simulations, we performed a numerical test with our code, without the planet and with an initially totally Keplerian distribution of velocities. The aim of this test is to analyse the survival time of the Keplerian kinematics before consistent radial perturbations comes out.
\begin{figure}
\resizebox{\hsize}{!}{\rotatebox{-90}{\includegraphics[clip=true]{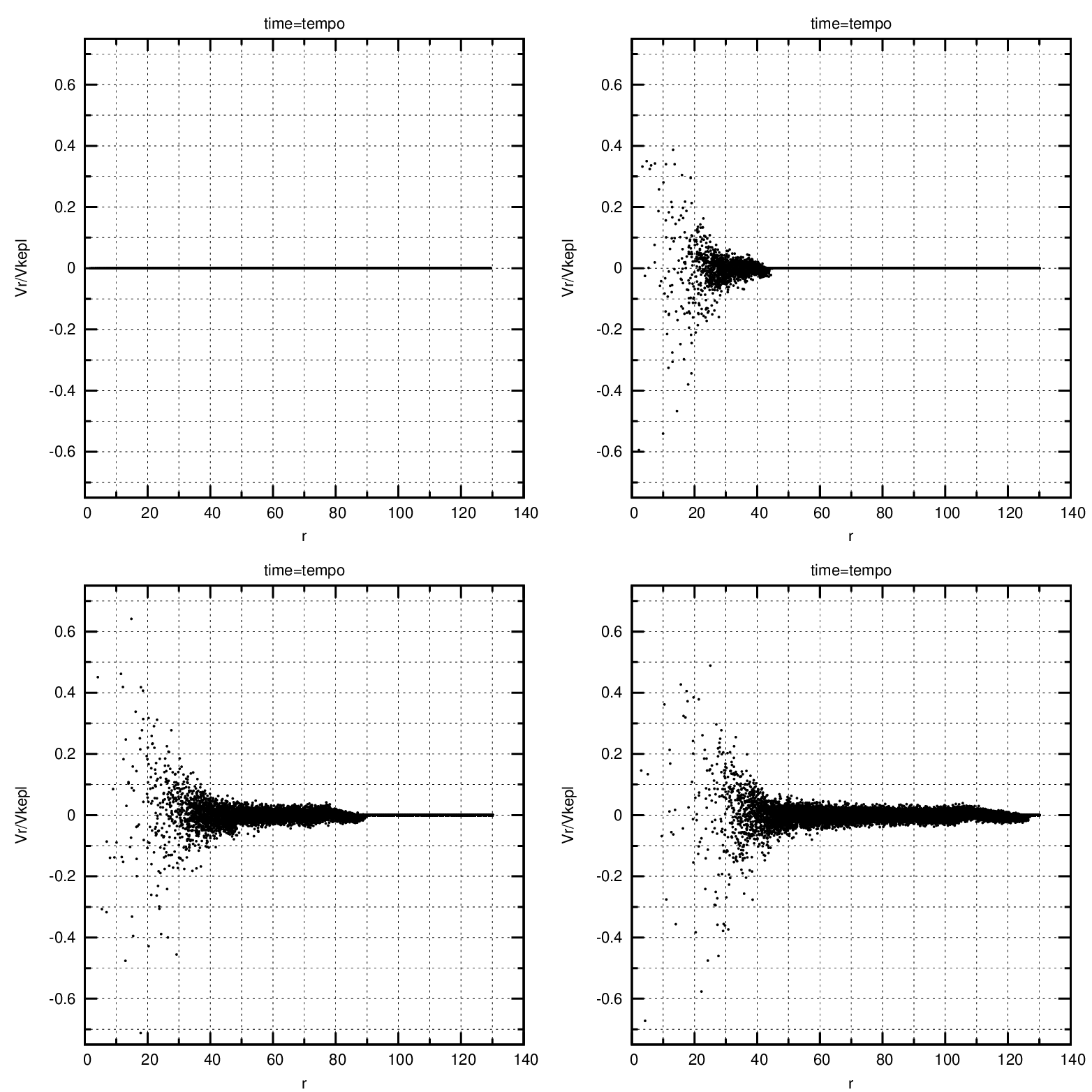}}}
 \caption{Distribution of the ratio ``radial speed/Keplerian speed'' as a function of radial distance for an initially totally Keplerian model. The four graphs show snapshots obtained at various times, considering that $time=1000$ is one orbital period at a distance $r=100$.}
\label{keplerian_breaking}
\end{figure}
Results are shown in figure \ref{keplerian_breaking}, where the ratio ``radial speed/Keplerian speed'' is plotted as a function of the radial distance to the central star. Four snapshots are provided at different times, the first at time zero, and the last at $time=3000$ (corresponding to three Keplerian orbits at $r=100$). A spontaneous breaking of the initially Keplerian kinemetics is observed, due probably to the contribution of pressure forces and to the collisional nature of our model.
\section{Discussion}
\label{discussion}
\begin{figure}
 \resizebox{\hsize}{!}{\rotatebox{-90}{\includegraphics[clip=true]{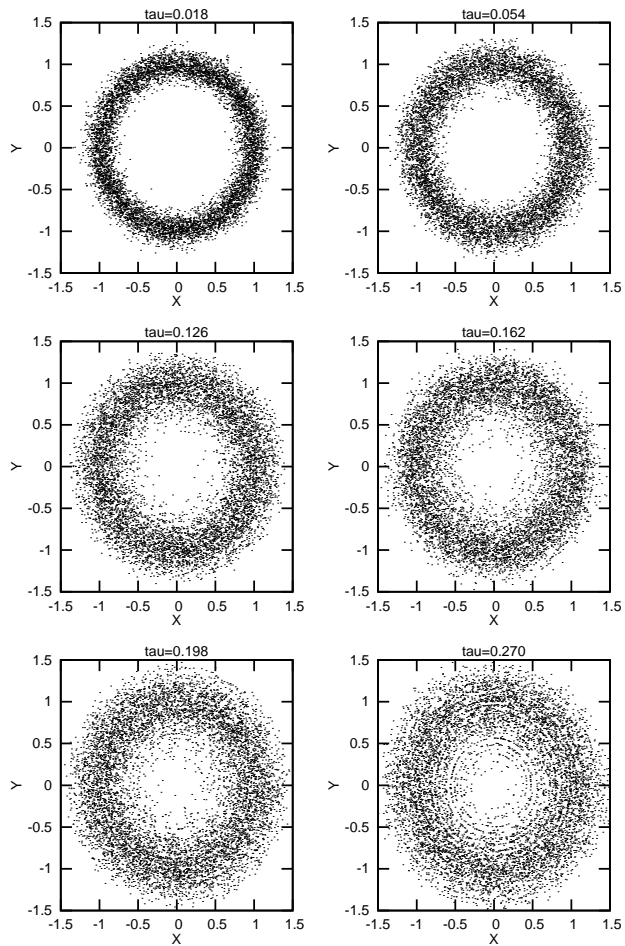}}}
 \caption{Evolution of a ring of particles with an initial distribution of 40000 particles as specified in \citet{Speith03}. $X$ and $Y$ positions are in units of $R_0$ and snapshots at various $\tau$ values are shown (see text).}
\label{ring_test}
\end{figure}

In a standard disc model, the relation connecting the disc thickness H, the disc radius r is: $H/r = c_s/v_{kepl}$, where $c_s$ is the sound velocity and $v_{kepl}$ is the Keplerian velocity at the disc outer edge. Such relation correlates the disc geometry to a tight Keplerian kinematic condition.
In our Keplerian disc model, the velocity of particles at the injectors positions is $v_{kepl}=0.55$ in the units specified in section \ref{models}, $c_s = 0.05$ being the sound speed, so that we obtain $H/r \simeq 0.05/0.55 ~ \sim 0.09$.
Generally this evaluation applies to a thin disc, where the hypothesis of vertical hydrostatic equilibrium is taken into account \citep{Terquem00}.
If this relation is translated also to sub-Keplerian modelling, it would give $H/r \simeq 0.12, 0.18, 0.36$ for $j=54,36,18$ respectively. However, although these results could give some indication, at least as a rough evaluation, these disc spreads are not strictly correct whenever turbulence and sub-Keplerian conditions are taken into account.
The results of this work deal with protoplanet migration in a protoplanetary accretion disc around a pre-main sequence star, according to the hypothesis that the accretion process is not strictly Keplerian. The Keplerian constraint is often used in computational models of accretion proto-planetary accretion discs on the basis of observational features that can be reproduced with Keplerian discs. It is sustainable if pressure and other forces are not able to significantly modify an initially Keplerian distribution of velocities.
We decided to explore the consequences of the presence of sub-Keplerian flows as a parametric study on the initial angular momentum, releasing the Keplerian constraint.\\
Our models are technically considered as physically inviscid, since they are based on the Euler equations. Artificial viscosity \citep{Monaghan92} is necessarily introduced in our models to resolve shocks numerically, to avoid spurious heating and to handle fluid discontinuities. Artificial viscosity vanishes when the limit value of the particle interpolation domain goes to zero. The linear $\alpha_{SPH}$ and quadratic $\beta_{SPH}$ artificial viscosity terms (usually $\sim 1$ and sometimes, in some specific cases, $<1$) are chosen to be 1 and 2, respectively.
\citet{Meglicki93,Drimmell96,Murray96,Okazaki02} demonstrated that the linear component of the artificial viscosity, in the continuum limit, yields a viscous shear force. \citet{Murray96} and \citet{Okazaki02} explicitly formulated such an artificial viscosity contribution in momentum and energy equations, founding an analogy between the shear viscosity generated by the linear artificial viscosity term and the well-known \citet{Shakura73} shear viscosity in the continuum limit. In the SPH method, the quadratic ($\beta_{SPH}$, von Neumann-Rychtmyer like viscosity) artificial viscosity term is used in order to handle structures.
An analytical formulation, describing the numerical artificial viscosity coefficient, is reported in \citet{Molteni91}: $\nu_{num} = c_s h$, where $c_s$ is the sound velocity.
Therefore, in our modelling, any radial transport process is due to the role of the artificial viscosity.
In order to estimate what $\alpha_{SS}$ value the numerical viscosity refers to, we can use the equality (see \citet{Lanzafame08} and references therein for a more detailed discussion of this point):
\begin{equation}
\nu= \alpha_{SS} c_s H = c_s h ~~~ \rightarrow ~~~ \alpha_{SS} = h/H
\label{alpha_ss}
\end{equation}
Using the above estimated $H/r$ values we obtain: $\alpha_{SS} \simeq 2.5\times 10^{-2}; ~~ 1.9\times 10^{-2}; ~~ 1.3\times 10^{-2}; ~~ 6.4\times 10^{-3}$ for Keplerian and $j=54,36,18$ respectively.\\
In order to allow a comparison between the effects produced by our implementation of artificial viscosity and the effects produced by physical shear viscosity, we performed a test comparing a simulated spreading ring \citep{Speith99, Speith03} with its approximate analytical solution. As in the above specified papers we considered an initial ring of $40000$ particles with a distribution of density given by the analytical solution obtained putting the dimensionless time $\tau=12 \nu t/R_0^2$ \citep{Speith03} equal to $0.018$. The initial ring radius $R_0$ is chosen to be equal to the stellar radius, while the kinematic viscosity is estimated as $\nu \sim c_s h$, where $h$ was set equal to $0.09$ as specified by \citet{Speith99}.\\
Graphs showing the results of this test, at various times (specified by $tau$ values) are shown in figure \ref{ring_test}, which can be directly compared to figure 1 in \citet{Speith03}.
Beware that this comparison, while allowing to show the shear effects produced by artificial viscosity, has its limits. In \citet{Speith03} the disc is totally pressure-less and viscosity is treated via the viscous stress tensor. In our case, even though we artificially removed the gas pressure, we could not remove the pressure contribution due to the artificial viscosity and we did not introduced a direct treatment of shear viscosity (our code is based on the Euler equations). Moreover, the relation \ref{alpha_ss} links the artificial viscosity and the kinematic viscosity to the value of the speed of sound $c_s$, so that the gas cannot be considered pressure-less.
The main difference between our test results and those obtained by \citet{Speith03} is the absence, in our case, of spiral structures.\\
Concerning the difference between the faster migration in our sub-Keplerian models and slower migration in Keplerian ones, we can notice that:
in sub-Keplerian models the most important contribution to downward migration seems to come from momentum transfer between the gas particles and the protoplanet; the sub-Keplerian $j_z$ value induces an average decrease of planet $l_z$, until the accretion disc centrifugal barrier is reached;
an initially Keplerian distribution of gas particles velocities, like in our Keplerian models, makes this effect less important, particularly with the Jovian planet. Models $kepl\_a$ and $kepl\_b$ show in fact a slow outward migration, despite the opening of a density gap.
This result is different from classical views on the subject, which connect the opening of a density gap to a viscosity driven downward migration (type II migration), but without physical viscosity (our models) the result can be different.\\
Some people \citep{Papaloizou06} suggest a type III migration mechanism, where the main migration ``engine'' is provided by material flowing through the co-orbital region. This material would change orbit (and angular momentum) while passing trough the co-orbital, exerting a torque onto the protoplanet. this torque can be evaluated trough momentum conservation.
This effect might actually be present in our models, but it's not the main mechanism, at least in our sub-Keplerian models. In our case, momentum conservation is directly applied to correct the protoplanet kinematics when disc particles are captured by the planet.
A rough estimate of the specific angular momentum change due to the capture of one disc particle can be done as follows: let $l$, $j$, $M$ and $m$ be the specific angular momentum and mass of the protoplanet and of the gas particle respectively; angular momentum conservation after particle capture leads to $\Delta l= m/(M+m) (j-l)~ \sim~ m/M (j-l)$ for $m/M << 1$; in our case (planet at an initial distance of $100$ in a circular Keplerian orbit) $l_{initial}= 62.8$; referring to our $36c$ model we have $m/M= 10^{-4}$ and $j=36$, so that after a single capture $\Delta l\sim -2.7\times 10^{-3}$.
About $10^5$ particles are captured during our simulation in model $36c$ before the angular momentum is reduced to about one half of the initial value (not all of them will have preserved the $j=36$ before capture), so that this capture mechanism is perfectly able to account for our rapid migration.\\
In our Keplerian models particle captures can both increase or decrease the protoplanet angular momentum, and a flow of particles through the co-orbital region is visible (figure \ref{Keplerian_disc_Jovian}), so type III mechanism might play a role. Notice however that the above specified flowing particles are moving towards smaller orbits, so they are mostly loosing angular momentum. This might actually be the reason for our slowly increasing angular momentum of the protoplanet in $kepl\_a,b$ models. But, being particle captures present in all our models, we cannot have pure type III migration anyway.\par
The results also lead our attention to the secondary role of the presence of a ``pseudo-atmosphere'' surrounding the protoplanets with the aim to discover whether such cumulative role could be also relevant.\\
The impact of the eventual presence of a pseudo-atmosphere in the protoplanets is visible in some models, particularly those with the less massive protoplanets (look at the difference between models $54c$ and $54d$), but never appears crucial in determining the main results. The slightly faster migration rate shown by models without the pseudo-atmosphere was actually expected, because the pseudo-atmosphere adds a repulsive pressure interaction between the protoplanet and the gas, that should hinder gas particle capture by the protoplanet.
Since part of the migration effect in our models is due to momentum transfer between the infalling particles and the protoplanets when the former are captured, a lower capture rate might bring a lower migration rate.\\
Particularly interesting is the result reported by the $c,d$ models, where the radial migration breaks down close to $\sim 15; 20; 45$ stellar radii, for $j=18; 36; 54$ respectively.
We believe that this effect is mainly due to the interaction of the protoplanet with the propagation of outward gas waves coming from inner zones (figures \ref{wave_l18}-\ref{wave_l36}), where the centrifugal barrier is located in our sub-Keplerian models. These results show that, despite the sub-Keplerian and more realistic conditions, planetary migration could even halt. Subsequent disc extinction could freeze the protoplanet orbital parameters.
Of course, due to a much higher inertia, Jovian planets evolve much more slowly.\\
An analytical expression giving us an idea of the protoplanet evolution, as a function of the model initial conditions, is expressed by the specific angular momentum fit equations \ref{exp_template}-\ref{linear_template} and by their temporal first derivatives.\\
Even if it is not the main topic of this paper, let's draw some comments on eccentricity behaviour shown in the bottom rectangles of figures \ref{l18_a_b}-\ref{Keplerian_c_d}. For Jupiter-like planets the orbits remain almost circular, with values around $10^{-5}$, while for Earth-like planets some eccentricity enhancing effect is visible in sub-Keplerian models, but the behaviour is quite complex and a more stable value is obtained in late migration phases, with values in the $0.01-0.02$ range.\\
For a comparison with other results we refer to the study performed by \citet{Shafer04} since they used SPH to simulate accretion disc hydrodynamics, even if some differences have to be remarked: first, they use a different numerical method to avoid particle mutual penetration, since they rely on XSPH with a bulk viscosity contribution to the viscous stress tensor; second, they simulate physical viscous discs according to \citet{Shakura73} prescription whilst our discs are inviscid; third, they use an isothermal equation of state whilst we rely on a classical ideal gas equation with variable thermal energy per unit mass; fourth, their initial velocity distribution is Keplerian.
They estimate a decrease of the planet distance from the the star of about 1 AU (from 5.2 AU to 4.1 AU) in about 30000 years for a Jovian planet, which brings to a rate of $Da/Dt \sim 3.3\times 10^{-5}~ AU/y$.
They also conclude that migration is braked by disc mass loss, which means a lowering of the disc density and of the gravitational interaction.\\
Inserting in our dimensionless model the following reference physical units:
central star mass $M_0 = 1 M_{\odot}$; stellar radius from typical T-Tauri stars \citep{Bertout89} $R_0 \sim 2 R_{\odot}$; the basic unit for time is $T_0 \sim 2.8\times 10^4 ~ \rm{s}$.
Our sub-Keplerian models with $j=18; 36; 54$ give then migration rates of $Da/Dt \sim 8\times 10^{-5}; 5.3\times ^{-5}; 3.4\times 10^{-5}~ AU/y$ respectively.
We notice that our $j=54$ (the less sub-Keplerian model) migration rate is similar to the average migration rate obtained by \citet{Shafer04}, even if our starting distance is about $1~AU$ with the above specified choices for physical units. More important, the braking effect is due, in our simulations, to a completely different mechanism based on turbulence, outward propagation of bouncing waves and changing of the average momentum transferred to the planet during gas particle captures, and we do not observe a significant lowering of the disc density.

\section*{Acknowledgements}
\label{acknowledgements}

The computational resources used to develop this work were provided by the COMETA consortium (http://www.consorzio-cometa.it), through the PI2S2 project (http://www.pi2s2.it). A special thank goes to Alessandro Grillo for useful help and technical support in using the COMETA computational GRID resources.

\label{lastpage}

\end{document}